\documentclass[11pt,twoside]{article}

\usepackage{asp2006}
\usepackage{psfig}

\begin{document}

\def\MSUN{\rm M_{\odot}}
\def\RSUN{\rm R_{\odot}}
\def\MSUNYR{\rm M_{\odot}\,yr^{-1}}
\def\MDOT{\dot{M}}

\newbox\grsign \setbox\grsign=\hbox{$>$} \newdimen\grdimen \grdimen=\ht\grsign
\newbox\simlessbox \newbox\simgreatbox
\setbox\simgreatbox=\hbox{\raise.5ex\hbox{$>$}\llap
     {\lower.5ex\hbox{$\sim$}}}\ht1=\grdimen\dp1=0pt
\setbox\simlessbox=\hbox{\raise.5ex\hbox{$<$}\llap
     {\lower.5ex\hbox{$\sim$}}}\ht2=\grdimen\dp2=0pt
\def\simgreat{\mathrel{\copy\simgreatbox}}
\def\simless{\mathrel{\copy\simlessbox}}

\title{Theory of winds in AGNs}   
\author{Daniel Proga}   
\affil{Department of Physics, University of Nevada, Las Vegas, NV 89154, USA}    

\begin{abstract} 
I  present a brief review of theory of winds in active galactic 
nuclei (AGN). Magnetic, radiation, and thermal driving likely
operate in AGN. In many cases, it is difficult to distinguish, 
both from observational and theoretical 
point of view, which of these wind driving mechanisms dominates
in producing winds.
Therefore, I focus on specific 
theoretical predictions which could help to improve 
our understanding of the physics of AGN winds.

\end{abstract}



\section{Introduction}

Compactness, high luminosity, time variability, spectral energy distribution
and spectral features of active galactic nuclei (AGN) are best explained
as consequences of disk accretion onto a supermassive black hole (SMBH).
However, we do not have any direct observational evidence that gas indeed
accretes onto a SMBH. Instead we quite commonly observe the opposite, i.e.,
gas outflows in AGN.

Broad absorption lines (BALs) in ultraviolet (UV) spectra of quasars are 
the most spectacular manifestation for such outflows. 
They are almost always blueshifted relative to
the emission-line rest frame, 
indicating the presence 
of winds from the active nucleus, with velocities as large as 
0.2~c (e.g., Turnshek 1998; for a few examples of BAL quasars with 
both blueshifted 
and redshifted absorption see Hall et. at. 2002).
BALs are observed not only in the UV but also 
in other wavelengths. For example, Chartas, Brandt \& Gallagher (2003) 
discovered a very broad absorption line in the X-ray spectrum of 
PG~1115+80. There are also a few examples of BALs in optical spectra of quasars
(Hutchings et al. 2002; Aoki et al. 2006; Hall 2006).
Other evidence for AGN winds include narrow absorption lines (NALs). 
UV spectra of some quasars show NALs which are blueshifted by as much 
as $\sim$~50000~${\rm km~s^{-1}}$ (Hamann et al. 1997). 
NALs are found much more commonly in the UV spectra of Seyfert galaxies
than in spectra of quasars, but  in Seyfert galaxies the lines are blueshifted 
only by several 100~${\rm km~s^{-1}}$ (Crenshaw et al. 1999).  
As BALs, NALs are observed not only in
the UV but also in  the X-rays. For example, 
Kaastra et al. (2000) observed NALs due to highly ionized species
in a high-resolution X-ray observation of the Seyfert
galaxy NGC~5548 obtained by Chandra.
The prominent broad emission lines (BEL) in the UV from H~I, O~VI, N~V, C~IV,
and Si~IV are the defining feature of  quasars (Blandford et al. 1990; 
Osterbrock 1989), and they may also be associated with a high
velocity wind (Murray et al. 1995, hereafter MCGV).

Many observations show that AGN winds are very complex flows, 
and neither spherical nor axial symmetry is applicable. 
For example, observations taken
with HST show bright emission-line knots in the narrow-line regions (NLRs) 
of some Seyfert galaxies which demonstrate a lack of any symmetry 
(Ruiz et al. 2005). Detailed analysis of these knots indicates that they 
may be related to the NLR clouds
outflowing from the nucleus, like the UV absorbers (Crenshaw \& Kraemer 2005).
Complex flow geometry is likely responsible for several distinct
components of absorption (e.g., NGC~3783, see Gabel et al. 2003; Mrk 279,
see Scott et al. 2004) and for transverse motion of absorbing gas seen
in one of the flow components of NGC~3783 (Crenshaw et al. 2003).
Additionally, Arav et al. (2005, and references therein) argued that to explain
the absorption in  the outflow observed in Mrk 279,
the absorber should be inhomogeneous. 

Generally, we would like to know the source
of the wind and how it is powered. Additionally, we would like
to know the wind geometry, energetics, content, and ionization state and what controls them.
Results obtained from data interpretation are very important but they also
show that without a physical model, data interpretation provides limited
insight into the wind origin and properties. 
To make progress in
understanding AGN winds, we need a physical multidimensional model
which will be capable of capturing the complex
wind geometry and dynamics and ultimately fit observations.

As mentioned above, many observational aspects of AGN can be explained
as consequences of disk accretion onto a SMBH. Therefore, I will focus
here on models which assume that
winds are also consequences of disk accretion. Specifically,
I will review models where an accretion disk is the source of
the wind and its energy. A much broader review and discussion
of other possibilities can be found for example in Krolik (1999)
and Crenshaw, Kraemer, \& George (2002).

\section{Driving Mechanisms}

The structure, dynamics and evolution of a wind can be 
described by the equations of radiation magnetohydrodynamics.
Possible wind driving mechanisms can be identified 
in the equation of motion:
\begin{equation}
   \rho \frac{D{\bf v}}{Dt} + \rho {\bf \nabla} \Phi =
- {\bf \nabla P} 
+ \frac{1}{4\pi} {\bf (\nabla \times B) \times B}
 + \rho {\bf F}^{rad},
\end{equation}
where $\rho$, ${\bf v}$, $P$, and ${\bf B}$  is the gas mass
density, velocity, gas pressure, and magnetic field,
respectively.
The terms ${\bf F}^{rad}$ and $\Phi$
in the equation of motion are
the total radiation force per unit mass and the gravitational potential, 
respectively.

To produce an outflow, there must be a force or forces which can 
overcome gravity. 
The terms in the right hand side of eq. 1 correspond to three possible forces: 
(1) the gradient of gas pressure, (2) the Lorentz force, and
(3) the radiation force (which results from gradient of radiation pressure).
Each of these forces can produce a wind. Here, I will 
describe how and under what physical conditions these forces
can be stronger than gravity. I consider
a Keplerian accretion disk, which is initially in a 
hydrostatic equilibrium (HSE), i.e., 
near the equatorial plane the gradient of gas pressure
balances gravity in the vertical direction.

\subsection{Thermal Driving}

A disk in HSE can lose its mass if the upper atmosphere is
heated. In the context of an AGN disk, significant heating
can result from irradiation of the outer/cooler parts
of the disk by the radiation from the inner/hotter disk. 
Theoretical models predict that  heating, especially X-ray heating, 
can have significant effects on the gas dynamics in disks. 
X-rays tend to heat low density gas, through Compton heating, to 
a high temperature of $\sim 10^7$~K (the so-called Compton temperature). 
With such a high temperature, an upper disk atmosphere  is
expected to either puff up and form a static corona, or to produce
a thermal wind, depending on
whether the thermal velocity exceeds the local
escape velocity, $v_{\rm esc}$ (e.g., Begelman, McKee and Shields, 1982; 
Ostriker, McKee, \& Klein 1991; Woods et al. 1996;
Proga \& Kallman 2002). Because $v_{\rm esc}$ is a function of radius,
for a given temperature of the heated gas, the effects of thermal
driving depend to the radius, too. 
Generally, thermal
driving is effective at large radii where $v_{\rm esc}$ is small
and one expects such winds to be responsible for low velocity
X-ray absorption features in AGN spectra (e.g., Chelouche \& Netzer 2005). 

Thermal driving is likely less important for temperatures below $10^7$~K
because the other two forces can dominate. In particular, radiation
forces can be significant for the gas with the temperatures $<10^5$~K. 

\subsection{Radiation Pressure Driving}

In the case of a fully ionized gas,
radiation pressure is only due to electron scattering.
Therefore, the system luminosity would have to be higher than 
the Eddington luminosity, $L_{\rm Edd}$ if
radiation driving is to be dynamically important. 
However, the radiation force can overcome gravity even for 
sub-Eddington sources provided the gas opacity is higher than 
the electron scattering, e.g., for the gas temperature  $< 10^5$~K.
The gas opacity can be enhanced via bound-bound and bound-free
transitions.

I will illustrate radiation driving by describing the radiation
force on spectral lines (line force) because 
a line-driven disk wind
is the most promising hydrodynamical (HD) scenario for
AGN outflows.  The presence of the BALs themselves
strongly indicates that substantial momentum is transfered from
a powerful radiation field  to the gas.
A crucial clue to the origin of AGN outflows comes from the discovery
of line-locking in BAL QSO spectra (e.g., Foltz et al. 1987).
Weymann et al. (1991)  discovered that a composite spectrum
of their BAL QSO sample shows
a double trough in the C~IV $\lambda$1549 BAL, separated in velocity space
by the N~V $\lambda$ 1240-Ly$\alpha$ splitting $\sim 5900$~km~$\rm s^{-1}$
(see also Korista et al. 1993). In fact Arav \& Begelman
(1994, see also  Arav et al. 1995 and Arav 1996) showed
that an absorption hump in the  C~IV $\lambda$ BAL is  ``the ghost
of Ly$\alpha$'' due to modulation of the radiation force by the
strong emission in Ly$\alpha$.

Line force can be significant, provided that the gas is moderately ionized and 
can interact with the UV continuum through very many UV line transitions
(Castor, Abbott \& Klein 1975, CAK hereafter). 
For gas photoionized by the UV radiation in the optically thin case, 
the total line opacity compared to the electron scattering (the
so-called CAK force multiplier) can be
as high as $M_{\rm max}\sim 2000-4000$ (e.g., CAK; 
Owocki, Castor \& Rybicki 1988).
For highly ionized gas, line force is inefficient
because of a lower concentration of ions capable of providing UV line opacity
($M_{\rm max}$ drops to zero).

In the radiation disk wind scenario, a wind is launched  from the disk by the
local disk radiation at radii where the disk radiation is mostly in the UV
(Shlosman, Vitello \& Shaviv 1985; MCGV). 
Such a wind  is continuous and
has mass loss rate and velocity which are capable of explaining
the blueshifted absorption lines observed in many AGN, if the ionization
state is suitable. This wind scenario has the important feature that
it does not rely on difficult to measure  forces or fields -- such as magnetic fields -- 
for their motive power.

UV driven disk winds in AGN are motivated by analogy with winds from hot stars,
which have been explored in great detail (e.g., Lamers and Cassinelli 1999
and references therein). However, stars differ from disks 
in important ways, including
the role of rotation near a Keplerian disk, the non-uniform disk temperature
distribution, and the influence of the strong X-ray flux from the inner disk
and black hole (the X-rays can overionize the gas).  
The wind density and mass loss rate can be
estimated from the UV observations of AGN.  When compared with
the X-ray flux from the AGN, the density is low enough that the gas is
predicted to  be highly ionized.  If so, the opacity needed for efficient
driving and for line formation will be very small and line driving
would be negligible.
On the other hand, the column density in the radial direction
close to the disk surface are high enough that a portion of the wind can
be shielded from this ionization.  

Viable models for disk winds must account
self-consistently for the wind driving, ionization, and self shielding.
MCGV pointed this out and postulated the existence of `hitchhiking' gas
which is not driven by UV but which provides the shielding.
They demonstrated the role of this shield gas on 
a one-dimensional, time independent quasi-radial flow.
Proga, Stone \& Kallman (2000, PSK hereafter; see also 
Proga \& Kallman 2004, PK hereafter) were able to relax some of
the MCGV assumptions, and explored consequences of a radiation driven disk
wind model by performing two-dimensional, axisymmetric, time-dependent HD simulations.
In PSK's simulations, self-consistent HD was coupled with
radiative driving, ionization  balance and radiative
transfer.
The main results from the simulations is the fact that if one allows
the disk wind to be two-dimensional,
the shielding material can be self-consistently produced as
a simple consequence of the multi-dimensional nature of the problem.

I emphasize that, within the framework of line driving,
X-ray shielding is necessary for an observer to see AGN outflows
in UV absorption lines and for the outflows
to be accelerated.
Assuming a steady state radial outflow and imposing total momentum
conservation,
one can show that unshielded gas outflowing with high velocities
has too high a photoionization parameter
to produce UV lines and to be driven by the line force.
This result holds for an outflow
consisting of a continuous wind as well as for
an outflow consisting of  dense clouds. The latter
needs shielding from X-rays if the filling factor decreases
with increasing radius.
A need for some shielding of clouds
has been already hinted at in the literature (e.g., de Kool \& Begelman 1995).

To avoid overionization without shielding, 
a wind which is launched
by a force other than the line force (and other than thermal driving) 
must be
denser than the wind predicted by the line-driven wind models
by several orders of magnitude.
Thus if AGN outflows are unshielded
and launched from an accretion disk at relatively large distances from
the central object
(where the local line force due to the disk is negligible) then we
have to figure out not only what mechanism launches
this  dense outflow but also what mechanism accelerates
the outflow to velocities much  higher
than $v_{\rm esc}$ from the location of the launch
(e.g., Arav, Li \& Begelman 1994).
The result that shielding of the wind from the X-rays
is required regardless  of the filling factor of the line-driven
outflow is consistent with
the observed anti-correlation for QSOs between
the relative strength of the soft X-ray flux and the C~IV
absorption equivalent width (e.g., Brandt, Laor, \& Wills 2000).

Thus, the importance  of line driving in quasars and in some other AGN has been
demonstrated in specific studies of this type of systems.
However, based on physical arguments as well as
numerical simulations one can show that radiation driven disk winds
are produced when $L_{\rm UV}> L_{\rm Edd}/M_{\rm max}$, where $L_{\rm UV}$
is the system UV luminosity
(e.g., Proga et al. 1998, PSD hereafter).
Moreover, one can argue that in all accretion disks, with the UV luminosity, 
$L_{\rm UV}~>$~a few $10^{-4}~L_{\rm Edd}$, 
mass outflows have been observed (Proga 2002 and fig. 2 there). For 
example, accretion disks around: massive black holes, 
white dwarfs (respectively as in AGN and cataclysmic variable, with 
$L_{\rm UV} > 0.001~L_{\rm Edd}$) 
and low mass young 
stellar objects (as in FU Ori stars with $L_{\rm UV}~>$ 
a few  0.01~$L_{\rm Edd}$ ) 
show fast winds. Systems that have too low UV 
luminosities to drive a wind include accretion disks around neutron stars and 
low mass black holes as in low mass X-ray binaries and galactic black holes. 
These systems indeed do not show outflows similar to those observed in 
AGN, cataclysmic variables, 
and FU Ori (see Miller et al. 2006, for an example of a disk wind
in a X-ray binary which is likely magnetically driven). 

Radiation seems to drive an outflow mainly for most luminous
and massive AGN. 
PK have performed simulations of radiation driven disk winds
for a range of the disk luminosities, $L_D$ 
and black hole masses, $M_{BH}$. 
In all explored cases, the disk atmosphere can 'shield' itself  from external X-rays
so that the local disk radiation can launch gas off the disk
photosphere. However, this gas is not always accelerated
to become a wind. 
In particular, for 
$M_{BH}=10^8~\MSUN$ of a typical quasar and
$L_D > 0.3 L_{\rm Edd}$, a strong disk wind develops
whereas for $L_D \simless 0.3 L_{\rm Edd}$ there is no wind.
For a less luminous disk or stronger ionizing radiation that reduces
$M_{max}$, or both, the line force can still lift gas off the disk
but it fails to accelerate it.
Such a failed disk wind has been
found in simulations with but also without X-ray ionization (see PSD for
no X-ray cases, runs 1 and 6 there).
Additionally, the disk wind solution is sensitive to
$M_{BH}$: for a fixed ratio, $L_D/L_{\rm Edd}$
it is easier to produce a wind for
$M_{\rm BH} \simgreat 10^7~\MSUN$ than for
$M_{\rm BH} \simless 10^7~\MSUN$. 
However, when a wind develops, the synthetic
line profiles predicted by the radiation driven models are consistent
with BALs observed in quasars (PK).

The fact that radiation driving is successful 
only for high $M_{\rm BH}$ and $L_{\rm D}$ is an important result
and its implications need to be explored (e.g., Proga 2005). 
One of the implications is of course
that radiation driving is not enough to explain all AGN. In fact,
some studies show that not all AGN outflows appear to be driven by 
radiation (e.g., Chelouche \& Netzer 2005; Kraemer et al. 2006).
This brings us to the third and last force that is important
in driving winds: the Lorentz force. 

\subsection{Magnetic Driving}

Magnetic effects must be important in AGN, in fact,
they appear to be essential for the existence of all accretion disks.  
The magnetorotational instability (MRI, Balbus \& Hawley 1991)
has been shown to be  a very robust and universal mechanism to produce
turbulence and transport angular momentum in disks at all radii
(Balbus \& Hawley 1998; see also Ji et al. 2006). 
It is therefore likely that magnetic fields control mass accretion inside 
the disk and play a key role in producing a mass outflow from the disk.

There are two classes of magnetically driven winds:
(1) magnetocentrifugal winds 
where the dominant
contribution to the Lorentz force is the magnetic tension and
(2) magnetic pressure driven winds 
where the dominant
contribution to the Lorentz force is the magnetic pressure.

Blandford \& Payne (1982; see also Pelletier \& Pudritz 1992) showed that  the centrifugal force can  
drive a wind from  the disk if the poloidal component of the magnetic
field, ${\bf B_p}$ makes an angle of $> 30^o$ with respect to the normal to 
the disk surface. Generally, magnetocentrifugal disk winds 
require  the presence of a sufficiently
strong, large-scale, ordered magnetic field threading the disk 
with a poloidal component at least comparable to the toroidal magnetic
field, $|B_\phi/B_p| \simless 1$ (e.g., Cannizzo \& Pudritz 1988, 
Pelletier \& Pudritz 1992). Several groups have studied
numerically outflows using this mechanism
(e.g., Ustyugova et al. 1995, 1999; Romanova et al. 1997; 
Ouyed \& Pudritz 1997a, 1997b, 1999; Krasnopolsky, Li \& Blandford 1999; 
Kato, Kudoh \& Shibata 2002).
An important feature of  magnetocentrifugal winds is that they require 
some assistance  to flow freely and steadily from the surface of the 
disk, to pass through a slow magnetosonic surface 
(e.g., Blandford \& Payne 1982). Thus
magnetocentrifugal wind models predict the geometry
and kinematics of the winds but they do not predict the mass-loss rate
but rather assume it. 
Therefore these models can not be completely tested against observations.

Many favor magnetocentrifugal winds  as an explanation for 
the outflows in young stellar objects and in some AGN
(e.g., Blandford \& Payne 1982; Uchida \& Shibata 1985; Emmering, Blandford \& Shlosman 1992;
Contopoulos \& Lovelace 1994; 
Bottorff et al. 1997;  Bottorff, Korista \& Shlosman 2000).
Magnetic winds do not require radiation pressure and thus can be important
in low luminosity  systems such as young stellar objects and
systems where gas can be overionized by very strong radiation
as in AGN (e.g.,  K$\ddot{\rm o}$nigl  \& Kartje 1994; Ouyed \& Pudritz 1997;
Ustyugova et al. 1999; Krasnopolsky, Li \& Blandford 1999,
and references therein). In the context of AGN outflows,
models usually rely on the effects of 
magnetic fields as well as on line driving
(K\"onigl \& Kartje 1994, de Kool \& Begelman 1995, 
Everett, K\"onigl, \& Arav 2002; Proga 2003a; Everett 2005). 

For example, within the framework
proposed by de Kool and Begelman (1995, see also Everett 2005) 
a wind is launched from an outer cold disk. 
Such a wind  is made of dense clouds confined by the magnetic field, and
therefore does not require shielding.
The wind is accelerated  by the UV radiation emitted from the inner disk. 
The wind may be but does not have to be
equatorial and can produce spectral features
for all inclination angles provided there is at least one cloud
in the line of sight. 

One of the differences between magnetocentrifugal and radiation driven disk winds 
is that the former corotate with the disk, at least close to the disk,
whereas the latter do not.
A key parameter shaping the total line profile  made up of both scattered 
emission and absorption  is the ratio of the expansion velocity to the rotational 
velocity (e.g., Proga 2003b). 
Corotating winds, conserve  the gas 
specific angular velocity and therefore have a higher rotational
velocity than those which do not corotate and conserve  
the gas specific angular momentum.
The former also have  higher terminal velocities due to stronger
centrifugal force. Additionally in the wind where the specific angular velocity
is conserved the rotational velocity is comparable to the terminal
velocity
while in the wind where the specific angular momentum is conserved
the rotational velocity decreases asymptotically to zero with
increasing radius
and is therefore much lower than the terminal velocity (Proga 2000).
Thus we should be able to distinguish these two kinds of winds
based on their line profiles.
For example, highly rotating  winds should produce emission lines  much
broader than slowly rotating, expanding winds, if we see
the disk edge-on. Line absorption should also be changed that corotation.
One should hope that synthetic line profiles predicted by
models of magnetocentrifugal winds will soon be computed so that
we will be able to compare these two types of winds.

Wind corotation depends on the strength of $B_p$. For a relatively
weak $B_p$, a wind may corotate only very close to the disk and be
driven by the magnetic pressure instead of the magnetic tension.
In particular, the toroidal magnetic field can quickly build up inside the disk
due to the differential rotation of the disk so that $|B_\phi/B_p| \gg 1$.
In such a case, the magnetic pressure of the toroidal field can give rise 
to a self-starting wind (e.g., Uchida \& Shibata 1985; 
Pudritz \& Norman 1986; Shibata \& Uchida 1986; Stone \& Norman 1994;
Contopoulos 1995; Kudoh \& Shibata 1997; Ouyed \& Pudritz 1997b).

To our best knowledge, the first
numerical simulations of
the two-dimensional, time-dependent magnetohydrodynamical (MHD) structure
of  line-driven winds from luminous accretion disks 
were presented in Proga (2003a). In these simulations,
the disk was initially threaded
by a purely axial magnetic field. This study was focused on a generic
disk wind problem and did not include strong ionizing radiation.
This approach is relevant to  AGN outflows
because it addresses, for example, the problem of a wind
produced beyond a shielding region, i.e., the outer outflow. 
The simulations showed 
that the magnetic field very quickly starts deviating
from purely axial due to MRI.
This leads to fast growth of the
toroidal magnetic field as field lines wind up due to the disk rotation.
As a result the toroidal field dominates over the poloidal field
above the disk and the gradient of the former drives a slow and dense disk 
outflow, which conserves specific angular momentum of fluid. 
Depending on the strength of the magnetic field relative to the 
system luminosity the disk wind can be radiation- or MHD-driven.
The pure line-driven wind consists of a dense, slow outflow that is 
bounded on the polar side by a high-velocity  stream. The mass-loss rate is 
mostly due to the fast stream. As the magnetic field strength increases first 
the slow part of the flow is affected, namely it becomes  denser and slightly 
faster and begins to dominate the mass-loss rate. In very strong magnetic 
field or pure MHD cases, the wind consists of only a dense, slow outflow 
without the presence of the distinctive fast stream 
which is a very characteristic feature of
line-driven winds. Winds launched by 
the magnetic fields are likely to remain dominated by the fields downstream 
because of their relatively high densities. Line driving 
may not be able to change a dense MHD wind because the line force
strongly decreases with increasing density.

\section{Concluding Remarks}

The main conclusion from the theoretical studies of AGN winds
is that these winds can not be explained  by just one driving mechanism.
This may seem like a disappointing outcome  of
decades of work. On the other hand,
one should not be surprised because
the AGN radiation is produced by a large accretion disk
with the ratio between the outer to inner radius of $>10^4$.
The amount but also type of the radiation emitted
by such a disk depend on the radius and can not be 
explained by just one radiative process (e.g., thermal emission). 
If the same disk loses its mass, as I assumed here, one should
not expect just one mechanism, even if it is a favorite one,
to dominate.

Future work should continue to consider eq. 1 with all its terms.
More importantly however, it should result not only in finding wind 
solutions but also in synthetic spectra predicted by these solutions.

\acknowledgements 
This work is supported by NASA through grants 
and HST-AR-10305 and HST-AR-10680 
from the Space Telescope Science Institute, 
which is operated by the Association of Universities for Research 
in Astronomy, Inc., under NASA contract NAS5-26555.



\begin{thebibliography}{}

\bibitem{} Aoki, K., Iwata, I., Ohta, K., Ando, M., Akiyama, M., \& Tamura,
N. 2006, ApJ, 651, 84

\bibitem{} Arav, N. 1996, ApJ, 465, 617

\bibitem{} Arav N., \& Begelman, M.C. 1994, ApJ, 434, 479

\bibitem{} Arav, N., Kaastra, J., Kriss, G.A., Korista, K. T., 
Gabel, J., Proga, D. 2005, ApJ, 620, 665

\bibitem[Balbus \& Hawley(1991)]{BH91}
Balbus, S.A., \& Hawley, J.F. 1991, ApJ, 376, 214

\bibitem[Balbus \& Hawley(1998)]{BH98}
Balbus, S.A., \& Hawley, J.F. 1998, Rev. Mod. Phys., 70, 1

\bibitem{}Begelman, M.C.,  McKee, C.F., Shields, G.A. 1983, ApJ, 271, 70

\bibitem[]{446} Brandt, W.N., Laor, A., \& Wills, B. J. 2000, ApJ, 528, 637

\bibitem{}Blandford, R.D., Netzer H., Woltjer L.,  
Courvoisier T., \& Mayor M. 1990, Active Galactic Nuclei, (Berlin: Springer)

\bibitem[Blandford \& Payne(1982)]{BP}
Blandford, R.D., \& Payne, D.G. 1982, MNRAS, 199, 883

\bibitem[Bottorff et al.(1997)]{Bottorff97}Bottorff, M., Korista,
  K.T., Shlosman, I., Blandford, R.D. 1997, ApJ, 479, 200

\bibitem[Bottorff, Korista \& Shlosman(2000)]{Bottorff00}Bottorff,
  M.C., Korista, K.T. \& Shlosman, I. 2000, ApJ, 537, 134

\bibitem[Cannizzo \& Pudritz(1989)]{CP}
Cannizzo, J.K., \& Pudritz, R.E. 1988, ApJ, 327, 840

\bibitem[Castor, Abbott \& Klein(1975)CAK]{CAK}
 Castor, J.I., Abbott, D.C., \& Klein, R.I. 1975, ApJ, 195, 157 (CAK)

\bibitem[]{} Chartas, G., Brandt, W. N., \& Gallagher, S. C. 2003, ApJ, 595, 85

\bibitem[Chelouche \& Netzer(2005)]{CN05}Chelouche, D. \& Netzer,
  H. 2005, ApJ, 625, 95

\bibitem[Chelouche \& Netzer(2006)]{}Chelouche, D. \& Netzer, H. 
2006, ApJ, 633, 693

\bibitem[Contopoulos(1995)]{C95}
Contopoulos, J. 1995, ApJ, 450, 616

\bibitem[]{}Contopoulos, J. \&  Lovelace, R.V.E. 1994, ApJ, 429, 139

\bibitem{} Crenshaw, D. M. \& Kraemer, S. B. 2005, ApJ, 625, 680

\bibitem[]{} Crenshaw, D. M., Kraemer, S. B., Boggess, A., Maran, S. P., 
Mushotzky, R. F., \& Wu, C. C. 1999, ApJ, 516, 750

\bibitem{}Crenshaw, D.M., Kraemer, S.B. \& George I.M.
  Mass Outflow in Active Galactic Nuclei: New Perspectives,
  ASP Conference Proceedings, Vol. 255. Edited by D. M. Crenshaw,
   S. B. Kraemer, and I. M. George, San Francisco: Astronomical Society
  of the Pacific, 2002

\bibitem{} Crenshaw, D. M., et al. 2003, ApJ, 594, 116

\bibitem[]{}de Kool, M. \& Begelman,  M.C. 1995, ApJ, 455, 448

\bibitem{}Emmering, R.T.,  Blandford, R.D. \&Shlosman, I. 1992, ApJ, 385, 460

\bibitem[Everett, K\"onigl, \& Arav(2002)]{EKA02}Everett, J.E.,
  K\"onigl, A. \& Arav, N. 2002, ApJ, 569, 671

\bibitem[Everett(2005)]{E05}Everett, J.E. 2005, ApJ, 631, 689


\bibitem{} Foltz, C.B., Weymann, R.J., Morris, S.L., \& Turnshek, D.A. 1987,
  ApJ, 317, 450

\bibitem{} Gabel J.R., et al. 2003, ApJ. 583, 178

\bibitem{} Hall P.B., 2006, AJ, in press (astro-ph/0611922)

\bibitem[]{} Hamann, F., Barlow, T.A., Cohen, R.D., Junkkarinen, V., 
\& Burbidge, E.M. 1997, in ASP Conf. Ser. 128, 
Mass Ejection from Active Galactic Nuclei, eds., N. Arav, 
I. Shlosman, \& R. Weymann (San Francisco: ASP), p. 19

\bibitem{} Hutchings, J. B., Crenshaw, D. M., Kraemer, S. B., Gabel, J. R.,
Kaiser, M. E., Weistrop, D., \& Gull, T. R. 2002, AJ, 124, 2543

\bibitem{} Ji, H., Burin, M., Schartman, E. \& Goodman, J. 2006, Nature 444, 343

\bibitem[]{} Kaastra, J. S., Mewe, R., Liedahl, D. A., Komossa, S.,
 Brinkman, A. C., 2000, A\&A, 354L, 83

\bibitem{}Korista, K.T., Voit, G.M. Morris, S.L., \& Weymann R.J. 1993, ApJS, 88, 357

\bibitem[Kato, Kudoh \& Shibata(2002)]{KKS}
Kato, S.X., Kudoh, T., \& Shibata, K. 2002, ApJ, 565, 1035

\bibitem[K\"onigl(1993)]{K93}
K\"onigl A. 1993, in  {\it ``Astrophysical Jets''},
  ed. D.P. O'Dea (Cambridge: CUP), 239

\bibitem[K\"onigl \& Kartje(1994)]{KK94}K\"onigl, A. \& Kartje,
  J.F. 1994, ApJ, 434, 446

\bibitem{} Kraemer S.B. et al. 2006, ApJ, 633, 693

\bibitem[Krasnopolsky, Li, \& Blandford(1999)]{KLB}
Krasnopolsky, R., Li, Z.-Y., \& Blandford, R. 1999, ApJ, 526, 631

\bibitem{}  Krolik, J.H. 1999, Active galactic nuclei: 
from the central black hole to the galactic environment, 
Princeton, N.J.: Princeton University Press

\bibitem[Kudoh \& Shibata(1997)]{KS}
Kudoh, T., \& Shibata, K. 1997, ApJ, 474, 362

\bibitem{} Lamers, H.G.J.L.M., and Cassinelli, J., 1999,``Introduction to
stellar winds'', Cambridge ; New York : Cambridge University Press

\bibitem{} Miller, J.M., Raymond, J., Fabian, A., Steeghs, D., Homan, J., 
Reynolds, C., van der Klis, M., \& Wijnands, R. 2006, Nature, 441, 953

\bibitem[Murray et al.(1995)]{MCGV}
Murray, N., Chiang, J., Grossman, S.A., \& Voit, G.M. 1995, ApJ,
451, 498

\bibitem{}Osterbrock, D.E. 1989, Astrophysics of Gaseous Nebulae 
and Active Galactic Nuclei (Mill Valley: University Science Books)

\bibitem[]{}  Ostriker, E. C., McKee, C.F., \& Klein, R.I. 1991, ApJ, 377, 5930

\bibitem[Ouyed \& Pudritz(1997a)]{OP97a} Ouyed, R.,
\& Pudritz, R.E. 1997a, ApJ, 482, 717

\bibitem[Ouyed \& Pudritz(1997b)]{OP97b} Ouyed, R.,
\& Pudritz, R.E. 1997b, ApJ, 484, 794

\bibitem{}  Ouyed R., Pudritz R.E. 1999, MNRAS, 309, 233


\bibitem[]{} Owocki, S. P., Castor, J. I., \& Rybicki, G. B. 1988, ApJ, 335, 914


\bibitem[Pauldrach, Puls, \& Kudritzki (1986)]{PPK}
Pauldrach, A., Puls, J., \& Kudritzki, R.P. 1986, A\&A, 164, 86

\bibitem[Pelletier \& Pudritz (1992)]{PP}  Pelletier, G.,
\& Pudritz, R.E. 1992, ApJ, 394, 117


\bibitem{} Proga, D. 2000, ApJ, 538, 684

\bibitem[Proga(2003a)]{P03a}  Proga, D. 2003a, ApJ, 585, 406

\bibitem[Proga(2003b)]{P03b}  Proga, D. 2003b, ApJ, 592, L9

\bibitem[Proga(2005)]{P05}  Proga, D. 2005, ApJ, 630, L9


\bibitem[Proga(2002)]{P02} Proga, D. 2002,
in Mass Outflow in Active Galactic Nuclei: New Perspectives, ASP
Conf. Proc. Vol. 255, ed. D.M. Crenshaw, S.B. Kraemer, \& I.M.
George (San Francisco: ASP), 309

\bibitem[]{} Proga, D., \& Kallman, T.R. 2002, ApJ, 565, 455 (PK)


\bibitem[Proga et al.(2002)]{Petal02}  Proga, D., Kallman, T.R.,
Drew, J.E., \& Hartley, L.E. 2002, ApJ, 572,  382

\bibitem[Proga, Stone, \& Drew(1998)PSD~98]{PSD98}  Proga, D., Stone,
J.M., \& Drew, J.E. 1998, MNRAS, 295, 595 (PSD)


\bibitem{} Proga, D., Stone J.M., \& Kallman T.R.\ 2000, ApJ, 543, 696 (PSK)

\bibitem[Pudritz \& Norman(1986)]{PN}
      Pudritz, R.E., \& Norman, C.A. 1986, ApJ, 301, 571

\bibitem{} Romanova, M.M., Ustyugova, G.V., Koldoba, A.V., Chechetkin, V.M., \&
  Lovelace, R.V.E. 1997, ApJ, 482, 708


\bibitem{}Ruiz, J.R., Crenshaw, D.M.,
  Kraemer, S.B., Bower, G.A., Gull, T.R., Hutchings, J.B., Kaiser,
  M.E., Weistrop, D. 2005, AJ, 129, 73

\bibitem{} Scott, J.E., et al. 2004, 152, 1

\bibitem{}  Shibata K.,  Uchida Y. 1986, PASJ, 38, 631

\bibitem[]{}Shlosman, I.,
  Vitello, P.A. \& Shaviv, G. 1985, ApJ, 294, 96

\bibitem[Stone \& Norman(1992)]{SN92} Stone, J.M., \&
Norman, M.L. 1992, ApJS, 80, 753

\bibitem[Stone \& Norman(1994)]{SN94} Stone, J.M., \&
Norman, M.L. 1994, ApJ, 433, 746

\bibitem[]{} Turnshek, D. A. 1988, in QSO Absorption Lines: Probing the Universe,
ed. J. C. Blades, D. A. Turnshek, \& C. A. Norman
(Cambridge: Cambridge Univ. Press), 17


\bibitem[Uchida \& Shibata(1985)]{US} Uchida, Y., \&
Shibata, K. 1985, PASJ, 37, 515

\bibitem{}  Ustyugova, G.V., Koldoba, A.V.,
Romanova, M.M., Chechetkin, V.M., \& Lovelace, R.V.E. 1995, ApJ,
439, 39L

\bibitem{}  Ustyugova G.V., Koldoba A.V., Romanova M.M. Chechetkin V.M. Lovelace R.V.E.
  1999, ApJ, 516, 221

\bibitem{} Weymann, R.J., Morris, S.L., Foltz, C.B., \& Hewett, P.C. 1991, ApJ, 373, 23


\bibitem[]{} Woods, D.T., Klein, R.I., Castor, J.I., McKee, C.F., \&
Bell, J.B. 1996, ApJ, 461, 767

\end{thebibliography}
\end{document}